\documentclass[a4paper]{article}

\usepackage{amsmath}
\usepackage{amssymb}
\usepackage{amstext}
\usepackage{graphicx}
\usepackage{fullpage}
\usepackage{xcolor}

\usepackage[skins,theorems]{tcolorbox}
\usepackage{authblk}

\DeclareMathOperator{\sign}{sgn}

\providecommand{\keywords}[1]
{
  \small	
  \textbf{\textit{Keywords---}} #1
}

\begin{document}

\title{A Finite Element Approach to the Numerical Solutions of Leland's Model}

\author[1,*]{Dongming Wei}
\author[2,*]{Yogi Ahmad Erlangga}
\author[1]{Gulzat Zhumakhanova}
\affil[1]{Nazarbayev University, Department of Mathematics, School of Sciences and Humanities, 53 Kabanbay Batyr Ave, Nur-Sultan 010000, Kazakhstan}
\affil[2]{Zayed University, Department of Mathematics, College of Natural and Health Sciences, Abu Dhabi Campus, P.O. Box 144534, United Arab Emirates}
\affil[ ]{\textit{dongming.wei@nu.edu.kz, yogi.erlangga@zu.ac.ae, gulzat.zhumakhanova@nu.edu.kz }}
\affil[*]{Corresponding author}

\date{ }
\maketitle

\begin{abstract} In this paper, finite element method is applied to Leland's model for numerical simulation of option pricing with transaction costs. Spatial finite element models based on P1 and/or P2 elements are formulated  in combination with a Crank-Nicolson-type temporal scheme. The temporal scheme is implemented using the Rannacher approach. Examples with  several sets of parameter values are presented and compared with finite difference results in the literature. Spatial-temporal mesh-size ratios are observed for controlling the stability of our method.  Our results compare favorably with the finite difference results in the literature for the model.
\end{abstract}


\section{Introduction} \label{sec:intro}

A fair option price in a complete financial market with no transaction costs can be modelled by the Black-Scholes equation~\cite{scholesB73, merton73}. The underlying assumption requires, however, that portfolio hedging takes place continuously. In the market with transaction costs, this assumption becomes unrealistically expensive. Modifications to the Black-Scholes model have been proposed to count for the transaction costs, which lead to various nonlinear models~\cite{leland85, Barles98S, Kratka98, Frey97S, Janda05S}.

In~\cite{leland85}, Leland proposes a modification to the Black-Scholes equation by allowing  portfolio rebalancing at a discrete time $\delta t$ with the transaction costs proportional to the value of the underlying asset. For an European call option with the strike price $K$ and expiration time $T$, its price $V$ at any time $t$ can then be modelled by using the nonlinear partial differential equation:
\begin{eqnarray}
V_t + \frac{1}{2}\sigma^2 S^2 (1 + Le \sign(V_{SS}))V_{SS} + r SV_S - r V = 0, \quad \text{in } (0,T) \times \mathbb{R}_+ \label{BSleland}
\end{eqnarray}
where $V = V(S,t)$, $S$ is the value of the underlying asset, $\displaystyle Le = \sqrt{\frac{2}{\pi}} \frac{c}{\sigma \sqrt{\delta t}}$ is the Leland number,  $r$ is the risk-free interest rate, $c$ is the round trip of the transaction cost per currency unit, and $\sigma$ is the volatility. For~\eqref{BSleland}, in addition to the boundary conditions
\begin{eqnarray}
  V(0,t) &=& 0, \label{bcond0} \\
  V(S,t) &=& S - K, \text { as } S \to \infty, \label{bcond1} 
\end{eqnarray}
for all $t \in [0,T]$, the following terminal condition at the expiration time $T$ is also required:
\begin{eqnarray}
   V(S,T) = \max(S-K,0). \label{icondT}
\end{eqnarray}
The condition~\eqref{icondT} is referred to as the pay-off function.

A few remarks are in order. Firstly, without transaction cost, $V_{SS} > 0$~\cite{scholesB73}. If this condition on $V_{SS}$ also holds in the nonlinear case, then we have
$$
  \tilde{\sigma}^2 := \frac{1}{2}\sigma^2 (1 + Le \sign(V_{SS})) = \frac{1}{2}\sigma^2 (1 + Le ) ;
$$
Thus ~\eqref{BSleland} is reduced to the Black-Scholes equation with an adjusted constant volatility $\tilde{\sigma} > \sigma$. Secondly, the Leland number depends highly on the transaction cost, which is typically a small percentage of the value of the assets and the time between rehedging, which is much smaller than the time to expiration. A small Leland number, usually when $Le < 1$, corresponds to a small transaction cost or large time interval between rehedging, which is considered high risk. In this case, the nonlinear terminal-boundary value problem~\eqref{BSleland} is known to be well-posed and has a unique solution $V(S,T)$ for any terminal condition  ~\cite{Avel94P}. See also, e.g., ~\cite{mm,dm} for mathematical analysis on the solution of~\eqref{BSleland}.

Practical option pricing is typically done by solving the underlying terminal-boundary value problem numerically. Popular numerical methods for this purpose are based on finite difference methods (FDM) and finite element (FEM) methods~\cite{Achdou05P}. FDM are particularly popular for both linear and nonlinear cases due to the simplicity of the methods, especially when the computation is performed on a uniform mesh~\cite{AnkudinovaE08,Company09JP, Zhao16YW}. Development of high-order methods as well as mesh adaptivity used to control numerical errors may however not be trivially done with FDM~\cite{Duering03FJ,liaoK09, Linde09PS, Gulen19PS}. These are not an issue with FEM, even though the implementation is more complex than FDM~\cite{Piron99H}. While FEM have been demonstrated to be a viable alternative to FDM in the linear cases~\cite{Marko08, Anda11AS, Golba13BA}, only limited work is presently done on the nonlinear cases, especially involving transaction costs under Leland's model (see~\cite{Almeida17CD}).

Our aim with this paper is to present some effective finite element methods for the Leland model and to demonstrate that the finite element method is also a practical choice for numerical simulations of problems in computational finance with Leland's model. 

We note here that since the initial work of Leland's, several authors have proposed modifications to the original Leland model~\eqref{BSleland}, to better capture the hedging strategy under transaction costs. These include, to mention a few, the model of Boyle and Vorst~\cite{Boyle92V}, Hoggard~\cite{Hoggard94WW}, and Zhao and Ziemba~\cite{Zhao04Z}. Such a modification is reflected in the adjusted volatility, which, however, shares a nonlinear term in common: the signum function $\sign(V_{SS})$. Along this line,~\cite{Avel94P} notes that, for $Le \ge 1$, Leland's model~\eqref{BSleland} is ill-posed under nonconvex terminal conditions and proposes a hedging strategy that fixes this issue. Despite all of these variants, in this paper, we shall focus only on the application of the finite element method for the original Leland model~\eqref{BSleland}. Application of the finite-element methods to the aforementioned variants can be easily done.

The remainder of the paper is organized as follows. After introducing transformation to Leland's model~\eqref{BSleland} in Section~\ref{LelandTrans}, we discuss a finite-element method and treatment for the signum term in Section~\ref{FEMleland}. Section~\ref{Timeintegrate} discusses the time-integration method. Numerical results are presented in Section~\ref{numerics}, followed by concluding remarks in Section~\ref{conclusion}.

\section{Leland's transformed model} \label{LelandTrans}

Following the standard strategy for solving the Black-Scholes-type problems, we first transform the terminal-boundary value problem~\eqref{BSleland} and \eqref{icondT} to an initial-boundary value problem of a simplified differential equation by using the following change of variables:
\begin{itemize}
  \item $\displaystyle \tau = \frac{1}{2} \sigma^2 (T - t)$ and hence $\displaystyle t = T - \frac{2 \tau}{\sigma^2}$;
  \item $\displaystyle x = \ln(S) + k \tau $ and hence $S = e^{x- k \tau}$; and
  \item $u(x,\tau) = e^{k \tau} V(S,t)$, and thus $V(S,t) = e^{-k \tau} u(x,\tau)$.
\end{itemize}
One can then show that
\begin{itemize}
   \item $\displaystyle V_t = \frac{\sigma^2}{2} e^{-k \tau} (-ku_x + ku - u_{\tau})$,
   \item $\displaystyle V_S = \frac{1}{S} e^{-k\tau} u_x$, and
   \item $\displaystyle V_{SS} = \frac{1}{S^2} e^{-k\tau}(u_{xx}-u_x)$.
\end{itemize}
Substitution of the above derivatives  of the option price $V$ into~\eqref{BSleland} yields the transformed Leland model
\begin{equation}
      u_{\tau} = u_{xx} -  u_x + Le |u_{xx} -  u_x|.  \label{Leland}
\end{equation}
Applying the above change of variables to the terminal and boundary conditions leads to
\begin{enumerate}
  \item the initial condition: $u(x,0) = \max(e^x - K,0)$,
  \item the boundary conditions
        \begin{itemize}
               \item $u(x,\tau) = 0$ as $x \to -\infty$
               \item $u(x,\tau) = e^x - K$ as $x \to \infty$.
        \end{itemize}
\end{enumerate}

As the problem is now defined in the unbounded spatial domain $(-\infty,+\infty)$, for computational purposes, we truncate the solution domain to $\Omega = [-R,R]$, where $0 < R < \infty$ and $R$ is taken to be a large number. We enforce the condition at $-\infty$ to be satisfied at $x = -R$, and similarly for the other boundary condition.

\section{Finite element methods} \label{FEMleland}

For the construction of the finite element approximation of the problem, we consider the following mixed formulation of~\eqref{Leland}:
\begin{eqnarray}
   &u_{\tau} = v + Le |v|, \label{Lelandu}\\
   &v = u_{xx} - u_x.  \label{Lelandv}
\end{eqnarray}
The Galerkin's finite element method for this formulation is based on the following  weak form of~\eqref{Lelandu} and~\eqref{Lelandv}. 

We begin by multiplying the equations~\eqref{Lelandu} and~\eqref{Lelandv} by the test functions $w$ and $z$, respectively, and integrate each over the domain $\Omega$ to get
\begin{eqnarray}
   \int\displaylimits_{\Omega} w (u_{\tau} -  (v + Le |v|) dx &=& 0, \notag \\
   \int\displaylimits_{\Omega} z (v - (u_{xx} - u_x)) d x &=& 0, \notag
\end{eqnarray}
which, after integration by parts, can be written as
\begin{eqnarray}
  \frac{\partial}{\partial\tau} \int\displaylimits_{\Omega} w u dx - \int\displaylimits_{\Omega} (wv + Le w |v|) dx &=& 0, \label{weak1} \\
   \int\displaylimits_{\Omega} z v dx  + \int\displaylimits_{\Omega} z_x u_x dx + \int\displaylimits_{\Omega} z u_x dx &=& 0.  \label{weak2}
\end{eqnarray}
Let $\displaystyle u = \sum_{i=1}^n u_i \psi_i + \sum_{i \in \mathcal{I}_{\partial \Omega}} u_i \psi_i$, $\mathcal{I}_{\partial \Omega} =\{0,n+1\}$, be the finite element approximation of the solution $u$,  where the second sum is the extension of the solution to the boundary $\partial \Omega=\{-R,R\}$ and $\psi_i$ is the global finite element shape function for the $i^{th}$ node in a spatial division $-R=x_0<...<x_i<...<x_{n+1}=R$. Similarly, we have  $\displaystyle v = \sum_{i=0}^{n+1} v_i \phi_i$, in which  no boundary conditions are set for $v$. Then \eqref{weak1} can be written as
\begin{eqnarray}
 0 &=& \frac{\partial}{\partial \tau} \int \displaylimits_{\Omega} w  \left\{  \sum_{i=1}^n u_i \psi_i + \sum_{i \in\mathcal{I}_{\partial \Omega}} u_i \psi_i\right\} d x - \int\displaylimits_{\Omega} (w\sum_{i=1}^n v_i \phi_i+ Le w |\sum_{i=0}^{n+1} v_i \phi_i|) dx \notag \\
   &=& \frac{\partial}{\partial \tau} \sum_{i=1}^n u_i \int\displaylimits_{\Omega} w \psi_i dx + \frac{\partial}{\partial \tau}  \sum_{i \in \mathcal{I}_{\partial \Omega}} u_i \int\displaylimits_{\Omega} w \psi_i dx - \sum_{i=1}^n v_i  \int \displaylimits_{\Omega} w \phi_i dx \notag \\
   &-& Le \int \displaylimits_{\Omega} w |\sum_{i = 0}^{n+1} v_i \phi_i |dx. \notag
\end{eqnarray}
Enforcing this condition to be satisfied by $n$ functions $w_j$, $j=1,\dots,n$ yields a system of $n$ equations
\begin{eqnarray}
0 &=& \frac{\partial}{\partial \tau} \sum_{i=1}^n u_i \int\displaylimits_{\Omega} w_j \psi_i dx + \frac{\partial}{\partial \tau}  \sum_{i \in \mathcal{I}_{\partial \Omega}} u_i \int\displaylimits_{\Omega} w_j \psi_i dx - \sum_{i=1}^n v_i  \int \displaylimits_{\Omega} w_j \phi_i dx \notag \\
   &-& Le \int \displaylimits_{\Omega} w_j |\sum_{i = 1}^n v_i \phi_i |dx. \label{int1}
\end{eqnarray}
For \eqref{weak2}, with $\displaystyle u_x = \sum_{i=1}^n u_i \psi_{i,x} + \sum_{i \in \mathcal{I}_{\partial \Omega}} u_i \psi_{i,x}$, we have
\begin{eqnarray}
   0 &=& \int\displaylimits_{\Omega} z \sum_{i=0}^{n+1} v_i \phi_i dx  + \int\displaylimits_{\Omega} z_x \left(\sum_{i=1}^n u_i \psi_{i,x} + \sum_{i \in \mathcal{I}_{\partial \Omega}} u_i \psi_{i,x} \right) dx \notag \\
     &+&  \int\displaylimits_{\Omega} z \left(\sum_{i=1}^n u_i \psi_{i,x} + \sum_{i \in \mathcal{I}_{\partial \Omega}} u_i \psi_{i,x}\right) dx \notag \\
     &=& \sum_{i=1}^n v_i  \int\displaylimits_{\Omega} z \phi_i dx +  \sum_{i=1}^n u_i \left\{ \int \displaylimits_{\Omega} z_x \psi_{i,x} dx + \int \displaylimits_{\Omega} z \psi_{i,x} dx \right\} \notag \\
    &+&  \sum_{i \in \mathcal{I}_{\partial \Omega}} u_i \left\{  \int \displaylimits_{\Omega} z_x \psi_{i,x} dx + \int \displaylimits_{\Omega} z \psi_{i,x} dx    \right\}.  \notag
\end{eqnarray}
By enforcing the above equation to be satisfied by $z_j$, $j = 0, \dots, n+1$ results in the system of $n$ equations
\begin{eqnarray}
0 &=& \sum_{i=0}^{n+1} v_i  \int\displaylimits_{\Omega} z_j \phi_i dx +  \sum_{i=1}^n u_i \left\{ \int \displaylimits_{\Omega} z_{j,x} \psi_{i,x} dx + \int \displaylimits_{\Omega} z_j \psi_{i,x} dx \right\} \notag \\
    &+&  \sum_{i \in \mathcal{I}_{\partial \Omega}} u_i \left\{  \int \displaylimits_{\Omega} z_{j,x} \psi_{i,x} dx + \int \displaylimits_{\Omega} z_j \psi_{i,x} dx    \right\}. \label{int2}
\end{eqnarray}

For our finite element models, we consider the Galerkin approach, where we set $w_i = z_i = \phi_i = \psi_i$. Equations~\eqref{int1} and~\eqref{int2} then become, for $j = 1,\dots,n$
\begin{eqnarray}
&&\frac{\partial}{\partial \tau} \sum_{i=1}^n u_i \int\displaylimits_{\Omega} \psi_j \psi_i dx + \frac{\partial}{\partial \tau}  \sum_{i \in \mathcal{I}_{\partial \Omega}} u_i \int\displaylimits_{\Omega} \psi_j \psi_i dx = \sum_{i=1}^n v_i  \int \displaylimits_{\Omega} \psi_j \psi_i dx  + Le \int \displaylimits_{\Omega} \psi_j |\sum_{i = 0}^{n+1} v_i \psi_i |dx. \label{eq12} \\
 &&\sum_{i=0}^{n+1} v_i  \int\displaylimits_{\Omega} \psi_j \psi_i dx =-\sum_{i=1}^n u_i \left\{ \int \displaylimits_{\Omega} \psi_{j,x} \psi_{i,x} dx +\int \displaylimits_{\Omega} \psi_j \psi_{i,x} dx \right\}-\sum_{i \in \mathcal{I}_{\partial \Omega}} u_i \left\{  \int \displaylimits_{\Omega} \psi_{j,x} \psi_{i,x} dx + \int \displaylimits_{\Omega} \psi_j \psi_{i,x} dx    \right\}.  \label{eq13}
\end{eqnarray}

Let the domain $\Omega$ be subdivided into $n_E$ nonoverlapping elements such that $\Omega = \bigcup\limits_{i=1}^{n+1} \Omega_{i}$, where $\Omega_{i}= [x_{i-1},x_i]$, the $i$-th element with boundary nodes $x_{i-1}$ and $x_i$. In this way, each integral above can be written as the sum of integral over each element. For instance
$$
   \int \displaylimits_\Omega \psi_j \psi_i dx = \sum_{\ell=1}^n \int \displaylimits_{\Omega_\ell} \psi_j \psi_i dx,
$$
and so on. Thus, the integral over the domain $\Omega$ can be evaluated by first evaluating integral over elements and then summing up, a process referred to as ``assembly''. In the implementation, the assembly process is based on element matrices that represents integral terms in~\eqref{eq12} and~\eqref{eq13} over each element $\Omega_j$. Structures of the element matrices depend on the specific choice of the functions $\psi_i$.
Specifically, the interpolation functions $\psi_i$ are chosen such that, at the nodal points $x_j$,
\begin{eqnarray}
    \psi_i(x_j) = \begin{cases}
                         1,& i = j, \\
                         0,&\text{otherwise}.
                    \end{cases}
\end{eqnarray}
In this way, at the left boundary point $x_0 = -R$,
$$
  u(x_0) = \sum_{i=n} u_i \psi(x_0) + \sum_{i \in \mathcal{I}_{\partial \Omega}} u_i \psi(x_0) = u_0 = 0.
$$
Similarly at the right boundary point $x_n = R$, 
$$
  u(x_n) = \sum_{i=n} u_i \psi(x_n) + \sum_{i \in \mathcal{I}_{\partial \Omega}} u_i \psi(x_n) = u_n = e^R - K.
$$

In the sequel, we discuss two interpolation functions used in our finite element methods.

\subsection{P1 finite element}

In the element $\Omega_j = [x_{j-1},x_j]$, with the meshsize $h_j = x_j-x_{j-1}$, we define two interpolation basis function:
\begin{eqnarray}
    \psi_{j-1}(x) &=& (x - x_j)/(x_{j-1} - x_j) = - (x-x_j)/h_j, \\
    \psi_j (x)      &=& (x-x_{j-1})/(x_j - x_{j-1}) = (x-x_{j-1})/h_j.
\end{eqnarray}
This is a linear (Lagrange) interpolation polynomial, which leads to the P1 (linear) finite element.

For the $\displaystyle \int \psi_j \psi_i dx$ term, the element matrix reads
\begin{equation*}
    M_j =  \begin{bmatrix}
                             \displaystyle  \int \displaylimits_{\Omega_{j}} \psi_{j-1} \psi_{j-1}dx & \displaystyle \int \displaylimits_{\Omega_{j}} \psi_{j-1} \psi_{j}dx \\
 \displaystyle \int \displaylimits_{\Omega_{j}} \psi_{j} \psi_{j-1}dx &\displaystyle  \int \displaylimits_{\Omega_{j}} \psi_{j} \psi_{j}dx
                           \end{bmatrix} = \frac{h_j}{6} \begin{bmatrix}
                                                         2 & 1 \\
                                                         1 & 2
                                                   \end{bmatrix}.
\end{equation*}
For the $-\displaystyle \int \psi_{j,x} \psi_{i,x} dx$ term, the element matrix reads
\begin{equation*}
    K_j =  -\begin{bmatrix}
                             \displaystyle  \int \displaylimits_{\Omega_{j}} \psi_{j-1,x} \psi_{j-1,x}dx & \displaystyle \int \displaylimits_{\Omega_{j}} \psi_{j-1,x} \psi_{j,x}dx \\
 \displaystyle \int \displaylimits_{\Omega_{j}} \psi_{j,x} \psi_{j-1,x}dx &\displaystyle  \int \displaylimits_{\Omega_{j}} \psi_{j,x} \psi_{j,x}dx
                           \end{bmatrix} = -\frac{1}{h_j} \begin{bmatrix}
                                                         1 & -1 \\
                                                        -1 & 1
                                                   \end{bmatrix}.
\end{equation*}
For the $\displaystyle \int \psi_{j} \psi_{i,x} dx$ term, the element matrix reads
\begin{equation*}
    P_j =  \begin{bmatrix}
                             \displaystyle  \int \displaylimits_{\Omega_{j}} \psi_{j-1} \psi_{j-1,x}dx & \displaystyle \int \displaylimits_{\Omega_{j}} \psi_{j-1} \psi_{j,x}dx \\
 \displaystyle \int \displaylimits_{\Omega_{j}} \psi_{j} \psi_{j-1,x}dx &\displaystyle  \int \displaylimits_{\Omega_{j}} \psi_{j} \psi_{j,x}dx
                           \end{bmatrix} = \frac{1}{2} \begin{bmatrix}
                                                         -1 & 1 \\
                                                         -1 & 1
                                                   \end{bmatrix}.
\end{equation*}
For the nonlinear term $\displaystyle \int \psi_{j} |\sum v_i \psi_i| dx$ term, since $\psi_j \ge 0$, $\psi_j \psi_i \ge 0$. Thus,
\begin{eqnarray}
   \int \displaylimits_\Omega \psi_j | \sum_{i=0}^{n+1} v_i \psi_i| dx &=& \int \displaylimits_\Omega  | \sum_{i=0}^{n+1} v_i \psi _j \psi_i| dx  \notag \\
&=& \int \displaylimits_\Omega | v_{j-1} \psi_j \psi_{j-1} + v_j \psi_j \psi_j + v_{j+1} \psi_j \psi_{j+1}|dx \notag \\
&=&  \int \displaylimits_{\Omega_j} | v_{j-1} \psi_j \psi_{j-1} + v_j \psi_j \psi_j + v_{j+1} \psi_j \psi_{j+1}|dx \notag \\
&+&  \int \displaylimits_{\Omega_{j+1}} | v_{j-1} \psi_j \psi_{j-1} + v_j \psi_j \psi_j + v_{j+1} \psi_j \psi_{j+1}|dx \notag \\
&=&   \int \displaylimits_{\Omega_{j}} | v_{j-1} \psi_j \psi_{j-1} + v_j \psi_j \psi_j |dx + \int \displaylimits_{\Omega_{j+1}} |  v_j \psi_j \psi_j + v_{j+1} \psi_j \psi_{j+1}|dx \notag \\
&\approx& |v_{j-1}|  \int \displaylimits_{\Omega_{j}} |\psi_j \psi_{j-1}|dx + |v_j| \left(  \int \displaylimits_{\Omega_{j}} |\psi_j \psi_j|dx +\int \displaylimits_{\Omega_{j+1}} |\psi_j \psi_j|dx  \right) \notag \\
&+& |v_{j+1}| \int \displaylimits_{\Omega_{j+1}} |\psi_j \psi_{j+1}|dx. \notag \\
&=&  |v_{j-1}|  \int \displaylimits_{\Omega_{j}} \psi_j \psi_{j-1}dx + |v_j| \left(  \int \displaylimits_{\Omega_{j}} \psi_j \psi_jdx +\int \displaylimits_{\Omega_{j+1}} \psi_j \psi_jdx  \right) \notag \\
&+& |v_{j+1}| \int \displaylimits_{\Omega_{j+1}} \psi_j \psi_{j+1}dx. \notag
\end{eqnarray}
The corresponding element matrix for the element $\Omega_{j}$ with nodal solution values $|v_{j-1}|$ and $|v_{j}|$ is given by
\begin{equation*}
    \bar{M}_j =  \begin{bmatrix}
                             \displaystyle  \int \displaylimits_{\Omega_{j}} \psi_{j-1} \psi_{j-1}dx & \displaystyle \int \displaylimits_{\Omega_{j}} \psi_{j-1} \psi_{j}dx \\
 \displaystyle \int \displaylimits_{\Omega_{j}} \psi_{j} \psi_{j-1}dx &\displaystyle  \int \displaylimits_{\Omega_{j}} \psi_{j} \psi_{j}dx
                           \end{bmatrix} = \frac{h_j}{6} \begin{bmatrix}
                                                         2 & 1 \\
                                                         1 & 2
                                                   \end{bmatrix}.
\end{equation*}
Notice that, for the P1 finite element, $\bar{M}_j = M_j$.

\subsection{P2 finite element}

In the basic element $\Omega_j$, we add a midpoint $x_{j-\frac{1}{2}}$, hence $x_j - x_{j-\frac{1}{2}} = h_j/2$, and define three quadratic interpolation polynomials
\begin{eqnarray}
    \psi_{j-1}(x) &=& \frac{(x-x_{j-\frac{1}{2}})(x-x_j)}{(x_{j-1} - x_{j-\frac{1}{2}})(x_{j-1} - x_{j})} =  2(x-x_{j-\frac{1}{2}})(x-x_j)/h_j^2, \\
   \psi_{j-\frac{1}{2}}(x) &=& \frac{(x-x_{j-1})(x-x_j)}{(x_{j-\frac{1}{2}} - x_{j-1})(x_{j-\frac{1}{2}} - x_{j})} = -4(x-x_{j-1})(x-x_j)/h_j^2, \\
    \psi_{j}(x) &=& \frac{(x-x_{j-1})(x-x_{j-\frac{1}{2}})}{(x_j - x_{j-1})(x_j - x_{j-\frac{1}{2}})} =  2(x-x_{j-1})(x-x_{j-\frac{1}{2}})/h_j^2.
\end{eqnarray}
The resulting finite element method is referred to as the P2 finite element.

For the $\displaystyle \int \psi_j \psi_i dx$ term, the element matrix reads
\begin{equation*}
    M_j =  \begin{bmatrix}
                             \displaystyle  \int \displaylimits_{\Omega_{j}} \psi_{j-1} \psi_{j-1}dx & \displaystyle \int \displaylimits_{\Omega_{j}} \psi_{j-1} \psi_{j-\frac{1}{2}}dx& \displaystyle \int \displaylimits_{\Omega_{j}} \psi_{j-1} \psi_{j}dx \\
 \displaystyle  \int \displaylimits_{\Omega_{j}} \psi_{j-\frac{1}{2}} \psi_{j-1}dx & \displaystyle \int \displaylimits_{\Omega_{j}} \psi_{j-\frac{1}{2}} \psi_{j-\frac{1}{2}}dx& \displaystyle \int \displaylimits_{\Omega_{j}} \psi_{j-\frac{1}{2}} \psi_{j}dx \\
 \displaystyle \int \displaylimits_{\Omega_{j}} \psi_{j} \psi_{j-1}dx  &\displaystyle  \int \displaylimits_{\Omega_{j}} \psi_{j} \psi_{j-\frac{1}{2}}dx &\displaystyle  \int \displaylimits_{\Omega_{j}} \psi_{j} \psi_{j}dx
                           \end{bmatrix} = \frac{h_j}{30} \begin{bmatrix}
                                                         4 & 2 & -1 \\
                                                         2 & 16 & 2 \\
                                                        -1 & 2  & 4
                                                   \end{bmatrix}.
\end{equation*}
For the $\displaystyle -\int \psi_{j,x} \psi_{i,x} dx$ term, the element matrix reads
\begin{equation*}
    K_j = - \begin{bmatrix}
                             \displaystyle  \int \displaylimits_{\Omega_{j}} \psi_{j-1,x} \psi_{j-1,x}dx & \displaystyle \int \displaylimits_{\Omega_{j}} \psi_{j-1,x} \psi_{j-\frac{1}{2},x}dx& \displaystyle \int \displaylimits_{\Omega_{j}} \psi_{j-1,x} \psi_{j,x}dx \\
 \displaystyle  \int \displaylimits_{\Omega_{j}} \psi_{j-\frac{1}{2},x} \psi_{j-1,x}dx & \displaystyle \int \displaylimits_{\Omega_{j}} \psi_{j-\frac{1}{2},x} \psi_{j-\frac{1}{2},x}dx& \displaystyle \int \displaylimits_{\Omega_{j}} \psi_{j-\frac{1}{2},x} \psi_{j,x}dx \\
 \displaystyle \int \displaylimits_{\Omega_{j}} \psi_{j,x} \psi_{j-1,x}dx  &\displaystyle  \int \displaylimits_{\Omega_{j}} \psi_{j,x} \psi_{j-\frac{1}{2},x}dx &\displaystyle  \int \displaylimits_{\Omega_{j}} \psi_{j,x} \psi_{j,x}dx
                           \end{bmatrix} = -\frac{1}{3h_j} \begin{bmatrix}
                                                         7 & -8 & 1 \\
                                                        -8 & 16 & -8 \\
                                                         1 & -8  & 7
                                                   \end{bmatrix}.
\end{equation*}
For the $\displaystyle \int \psi_{j} \psi_{i,x} dx$ term, the element matrix reads
\begin{equation*}
   P_j = \begin{bmatrix}
                             \displaystyle  \int \displaylimits_{\Omega_{j}} \psi_{j-1} \psi_{j-1,x}dx & \displaystyle \int \displaylimits_{\Omega_{j}} \psi_{j-1} \psi_{j-\frac{1}{2},x}dx& \displaystyle \int \displaylimits_{\Omega_{j}} \psi_{j-1} \psi_{j,x}dx \\
 \displaystyle  \int \displaylimits_{\Omega_{j}} \psi_{j-\frac{1}{2}} \psi_{j-1,x}dx & \displaystyle \int \displaylimits_{\Omega_{j}} \psi_{j-\frac{1}{2}} \psi_{j-\frac{1}{2},x}dx& \displaystyle \int \displaylimits_{\Omega_{j}} \psi_{j-\frac{1}{2}} \psi_{j,x}dx \\
 \displaystyle \int \displaylimits_{\Omega_{j}} \psi_{j} \psi_{j-1,x}dx  &\displaystyle  \int \displaylimits_{\Omega_{j}} \psi_{j} \psi_{j-\frac{1}{2},x}dx &\displaystyle  \int \displaylimits_{\Omega_{j}} \psi_{j} \psi_{j,x}dx
                           \end{bmatrix} = \frac{1}{6} \begin{bmatrix}
                                                         -3 & 4 & -1 \\
                                                        -4 & 0 & 4 \\
                                                         1 & -4  & 3
                                                   \end{bmatrix}.
\end{equation*}

To construct the element matrix corresponding to the $\displaystyle \int \psi_{j} |\sum v_i \psi_i| dx$ term, we begin with the splitting
$$
   \psi_j(x) = \psi_j^+(x) - |\psi_j^-(x)|
$$
where $\psi_j^+$ is the nonnegative part of $\psi_j$ and $\psi_j^-$ is the nonpositive part of $\psi_j$. Then
\begin{eqnarray}
   \int \displaylimits_\Omega \psi_j | \sum_{i=1}^n v_i \psi_i| dx &=& \int \displaylimits_\Omega (\psi_j^+ - |\psi_j^-|) | \sum_{i=1}^n v_i \psi_i| dx  \notag \\
&=&  \int \displaylimits_\Omega  | \sum_{i=1}^n v_i \psi_j^+ \psi_i| dx -  \int \displaylimits_\Omega  | \sum_{i=1}^n v_i |\psi_j^-| \psi_i| dx \notag \\
&\approx& \int \displaylimits_\Omega  \sum_{i=1}^n |v_i| |\psi_j^+ \psi_i| dx -  \int \displaylimits_\Omega  \sum_{i=1}^n |v_i| |\psi_j^- \psi_i| dx \notag \\
&=& \int \displaylimits_{\Omega_j} \left(  |v_{j-1}| |\psi_j^+ \psi_{j-1}| +  |v_{j-\frac{1}{2}}| |\psi_j^+ \psi_{j-\frac{1}{2}}| +  |v_{j}| |\psi_j^+ \psi_{j}| \right) dx \notag \\
&+& \int \displaylimits_{\Omega_{j+1}} \left(  |v_{j}| |\psi_j^+ \psi_{j}| +  |v_{j+\frac{1}{2}}| |\psi_j^+ \psi_{j+\frac{1}{2}}| +  |v_{j+1}| |\psi_j^+ \psi_{j+1}| \right) dx \notag \\
&-& \int \displaylimits_{\Omega_j} \left(  |v_{j-1}| |\psi_j^- \psi_{j-1}| +  |v_{j-\frac{1}{2}}| |\psi_j^- \psi_{j-\frac{1}{2}}| +  |v_{j}| |\psi_j^- \psi_{j}| \right) dx \notag \\
&-& \int \displaylimits_{\Omega_{j+1}} \left(  |v_{j}| |\psi_j^- \psi_{j}| +  |v_{j+\frac{1}{2}}| |\psi_j^- \psi_{j+\frac{1}{2}}| +  |v_{j+1}| |\psi_j^- \psi_{j+1}| \right) dx \notag \\
&=& |v_{j-1}|  \int \displaylimits_{\Omega_j} \left( |\psi_j^+ \psi_{j-1}| - |\psi_j^- \psi_{j-1}|\right) dx + |v_{j-\frac{1}{2}}|\int \displaylimits_{\Omega_j} \left(|\psi_j^+ \psi_{j-\frac{1}{2}}| -   |\psi_j^- \psi_{j-\frac{1}{2}}|\right) dx \notag \\
&+&  |v_{j}| \left\{ \int \displaylimits_{\Omega_j} \left(|\psi_j^+ \psi_{j}| - |\psi_j^- \psi_{j}|\right)dx + \int \displaylimits_{\Omega_{j+1}} \left(|\psi_j^+ \psi_{j}| - |\psi_j^- \psi_{j}|\right)dx\right\}  \notag \\
&+&   |v_{j+\frac{1}{2}}|  \int \displaylimits_{\Omega_j} \left( |\psi_j^+ \psi_{j+\frac{1}{2}}| - |\psi_j^- \psi_{j+\frac{1}{2}}|\right) dx + |v_{j+1}|\int \displaylimits_{\Omega_j} \left(|\psi_j^+ \psi_{j+1}| -   |\psi_j^- \psi_{j+1}|\right) dx. \notag
\end{eqnarray}

The above leads to the element matrix 
\begin{equation*}
   \bar{M}_j = \begin{bmatrix}
 \displaystyle  \int \displaylimits_{\Omega_j} \left( |\psi_{j-1}^+ \psi_{j-1}| - |\psi_{j-1}^- \psi_{j-1}|\right) dx &  \displaystyle \int \displaylimits_{\Omega_j} \left(|\psi_{j-1}^+ \psi_{j-\frac{1}{2}}| -   |\psi_{j-1}^- \psi_{j-\frac{1}{2}}|\right) dx &  \displaystyle \int \displaylimits_{\Omega_j} \left(|\psi_{j-1}^+ \psi_{j}| - |\psi_{j-1}^- \psi_{j}|\right)dx \\
 \displaystyle  \int \displaylimits_{\Omega_j}|\psi_{j-\frac{1}{2}}^+ \psi_{j-1}| dx &  \displaystyle \int \displaylimits_{\Omega_j} |\psi_{j-\frac{1}{2}}^+ \psi_{j-\frac{1}{2}}| dx &  \displaystyle \int \displaylimits_{\Omega_j} |\psi_{j-\frac{1}{2}}^+ \psi_{j}|dx \\
 \displaystyle  \int \displaylimits_{\Omega_j} \left( |\psi_j^+ \psi_{j-1}| - |\psi_j^- \psi_{j-1}|\right) dx &  \displaystyle \int \displaylimits_{\Omega_j} \left(|\psi_j^+ \psi_{j-\frac{1}{2}}| -   |\psi_j^- \psi_{j-\frac{1}{2}}|\right) dx &  \displaystyle \int \displaylimits_{\Omega_j} \left(|\psi_j^+ \psi_{j}| - |\psi_j^- \psi_{j}|\right)dx 
            \end{bmatrix}.
\end{equation*}
Note that for $\phi_{j-\frac{1}{2}}^- = 0$. Evaluating the integrals results in the element matrix
\begin{equation*}
   \bar{M}_j = \frac{h_j}{120}
                     \begin{bmatrix}
                          15 & 8 & 0 \\
                           8 & 64 & 8 \\
                           0 &  8  & 15
                     \end{bmatrix},
\end{equation*}
associated with the nodal solution $|v_{j-1}|$, $|v_{j-1/2}|$, and $|v_{j}|$. In this case, $\bar{M} \neq M$.

\section{The temporal integration scheme} \label{Timeintegrate}

The global finite element system can be assembled into the following first order ODE in matrix form
\begin{eqnarray}
  \frac{\partial}{\partial \tau} (M \mathbf{u} + \mathbf{b}_M) = M \mathbf{v} + Le \bar{M}|\mathbf{v}| := \mathbf{F},
\end{eqnarray}
where $\mathbf{v} = M^{-1}(K \mathbf{u} - P \mathbf{u} + \mathbf{b}_K - \mathbf{b}_P)$, and all boldface letters are used to denote the solution nodal value vectors or the finite element matrices. For an implementation of the assembly process, see, e.g., \cite{Prem04W}.

Integration over time is approximated using the Crank-Nicolson-type scheme:
$$
   \frac{1}{\Delta \tau} (M\mathbf{u}^{n+1} + \mathbf{b}_M^{n+1} - (M\mathbf{u}^{n} + \mathbf{b}_M^n)) = \theta \mathbf{F}^{n+1} + (1-\theta) \mathbf{F}^{n},
$$
with $\theta \in [0,1]$. $\theta = 0$ and 1 correspond to the explicit forward and implicit backward Euler method, respectively. Rearranging the above equation leads to
$$
   M\mathbf{u}^{n+1} - \theta  \Delta \tau \mathbf{F}^{n+1} =  M \mathbf{u}^n + (1-\theta) \Delta \tau \mathbf{F}^{n} - \mathbf{b}_M^{n+1} +  \mathbf{b}_M^n
$$
or
$$
   M\mathbf{u}^{n+1} - \theta  \Delta \tau (M \mathbf{v}^{n+1} + Le \bar{M}|\mathbf{v}^{n+1}|) =  M \mathbf{u}^n + (1-\theta) \Delta \tau \mathbf{F}^{n} - \mathbf{b}_M^{n+1} +  \mathbf{b}_M^n.
$$
With $\mathbf{v}^{n+1} = M^{-1}(K \mathbf{u}^{n+1} - P \mathbf{u}^{n+1} + \mathbf{b}_K^{n+1} - \mathbf{b}_P^{n+1}n)$, we get
$$
   M\mathbf{u}^{n+1} - \theta  \Delta \tau (K \mathbf{u}^{n+1} - P \mathbf{u}^{n+1} + Le \bar{M}|\mathbf{v}^{n+1}|) =  M \mathbf{u}^n + (1-\theta) \Delta \tau \mathbf{F}^{n} - \mathbf{b}_M^{n+1} +  \mathbf{b}_M^n + \theta \Delta \tau (\mathbf{b}_K^{n+1} - \mathbf{b}_P^{n+1}).
$$
This equation is nonlinear in the $(n+1)$ variables. Setting
$|\mathbf{v}^{n+1}| = |\mathbf{v}^n|$ results in the linearized form:
$$
   M\mathbf{u}^{n+1} - \theta  \Delta \tau (K \mathbf{u}^{n+1} - P \mathbf{u}^{n+1}) =  M \mathbf{u}^n + (1-\theta) \Delta \tau \mathbf{F}^{n} + \theta  \Delta \tau Le \bar{M}|\mathbf{v}^{n}| - \mathbf{b}_M^{n+1} +  \mathbf{b}_M^n + \theta \Delta \tau (\mathbf{b}_K^{n+1} - \mathbf{b}_P^{n+1})
$$
or
$$
   A \mathbf{u}^{n+1} =  M \mathbf{u}^n + (1-\theta) \Delta \tau \mathbf{F}^{n} + \theta  \Delta \tau Le \bar{M}|\mathbf{v}^{n}| - \mathbf{b}_M^{n+1} +  \mathbf{b}_M^n + \theta \Delta \tau (\mathbf{b}_K^{n+1} - \mathbf{b}_P^{n+1})
$$
with
\begin{eqnarray}
   A &=&  M - \theta  \Delta \tau (K - P), \notag \\
   \mathbf{F}^n &=& M \mathbf{v}^n + Le \bar{M}|\mathbf{v}^n|, \notag \\
   \mathbf{v}^n &=& M^{-1}(K \mathbf{u}^n - P \mathbf{u}^n + \mathbf{b}_K^n - \mathbf{b}_P^n) \notag
\end{eqnarray}

Recall that for the P2 FEM, $\bar{M} \neq M$. As another level of approximation, we may set also $\bar{M} = M$. We refer to the numerical method  described above as Version 1, and with the setting $\bar{M} = M$ is Version 2. For improved stability, the Crank-Nicolson method is implemented using the Rannacher approach~\cite{Rann84, Giles06C}, in which the first Crank-Nicolson step is replaced by a few backward implicit Euler steps with smaller time steps (e.g., $\Delta \tau_R = \Delta \tau/n_R$), where $n_R$ is the number of backward Euler time steps in from $\tau = 0$ to $\tau = \Delta \tau$).

\section{Numerical results and convergence} \label{numerics}

In this section, we present some numerical solutions computed using the FEM and the time-integration method discussed in Sections~\ref{FEMleland} and~\ref{Timeintegrate}. All results are computed on a uniform finite-element mesh, even though the method can be implemented on a nonuniform mesh.

\subsection{Solutions for the linear BS model}

In the absence of closed-form exact solution, to verify the accuracy of the methods, we first conducted a numerical test using a linear case, which corresponds to the situation where no rebalancing involves. This is equivalent to having $\delta t \to \infty$ and  therefore $Le = 0$. For this test, we set $r = 0.1$, $\sigma = 0.2$, the expiration time $T = 1$, and the strike price $K = 100$. 

\begin{figure}[!h]
\includegraphics[width=0.5\textwidth]{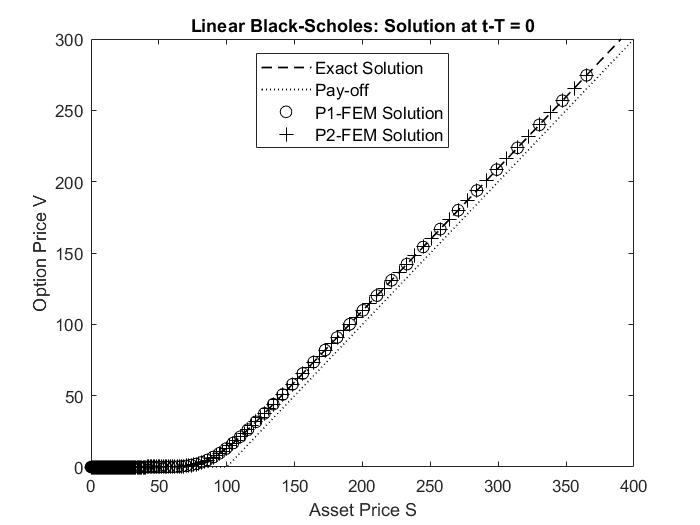}
\includegraphics[width=0.5\textwidth]{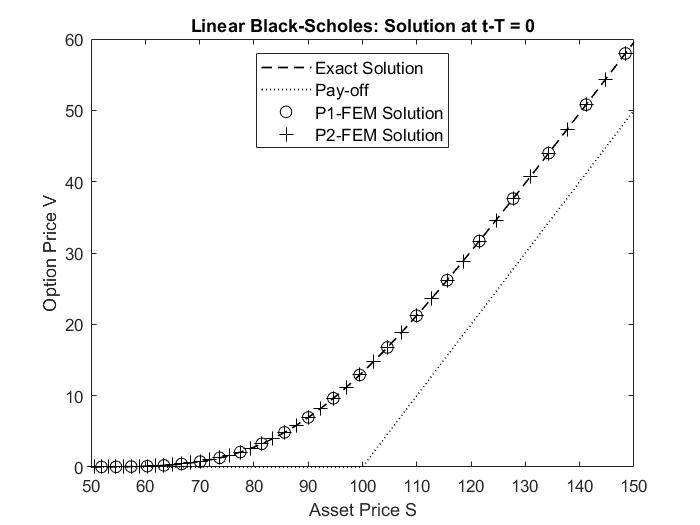}
\caption{Solution of the linear Black-Scholes equation}
 \label{fig:linear}
\end{figure}

The numerically computed option price $V$ is shown in Figure~\ref{fig:linear} as a function of asset price $S$. Comparing with the exact solution, we see in the figure the highly accurate solution obtained by the linear (P1)  and quadratic (P2) FEM. 

\subsection{Solutions for the (nonlinear) Leland model}

\begin{figure}[!h]
\includegraphics[width=0.5\textwidth]{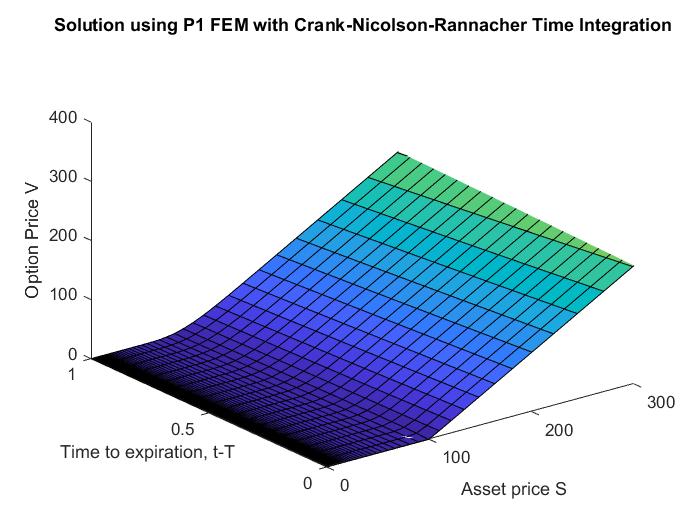}
\includegraphics[width=0.5\textwidth]{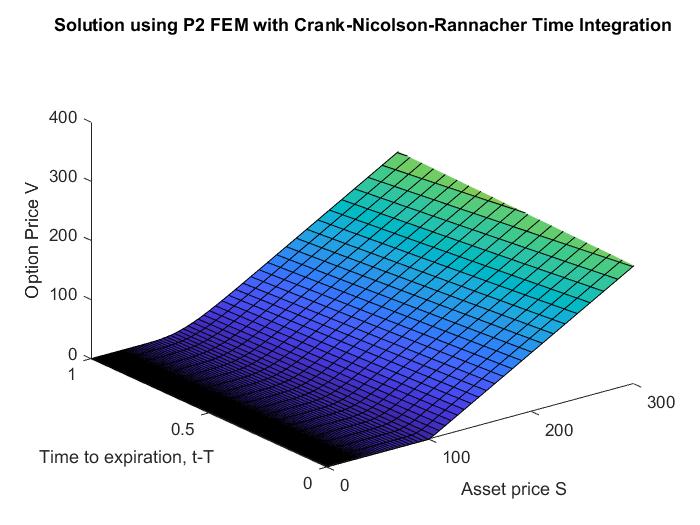}
\caption{Solution of Leland's  model over time $t \in [0,1]$, with $r = 0.1$, $\sigma = 0.2$, $K = 100$, $c = 0.01$, $\delta t = 0.01$, giving $Le \approx 0.4$;  Left figure: $h = 0.1$, $\Delta \tau = 0.001$ (thus, $\Delta \tau/h^2 = 0.1 \le 1$); Right figure: $h = 0.0125$, $\Delta \tau = 0.000125$ (thus, $\Delta \tau/h^2 = 0.8 \le 1$). In both cases, $\Delta \tau /h =  0.01$.} 
\label{fig:Le04complete}
\end{figure}

For the nonlinear Leland model, we consider cases, where (i) $Le < 1$, and (ii) $Le > 1$.
We discuss  the convergence and numerical stability of our schemes for numerical solutions of the two these cases in the following: 
\paragraph{The $Le < 1$ case.}
Our first test in this case is based on the parameters $r = 0.1$, $\sigma = 0.2$, $T = 1$, $K = 100$, $c = 0.01$, and $\delta t = 0.01$, yielding $Le = 0.4$. For the FEM, we set discretization parameters such that $\Delta \tau/h < 1$ and $\Delta \tau/h^2 < 1$, to control the stability of the method. Otherwise mentioned, we keep $\Delta \tau /h = 0.01$. The numerical solutions over time $t \in [0,1]$ are shown in Figure~\ref{fig:Le04complete}, for the P1 and P2 FEM. A look at the figure gives no indication of numerical instability.

\begin{figure}[!h]
\begin{center}
\includegraphics[width=0.45\textwidth]{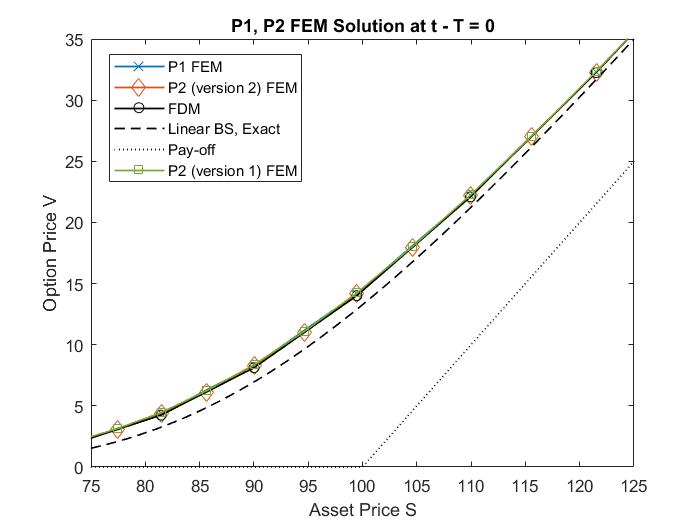}
\includegraphics[width=0.45\textwidth]{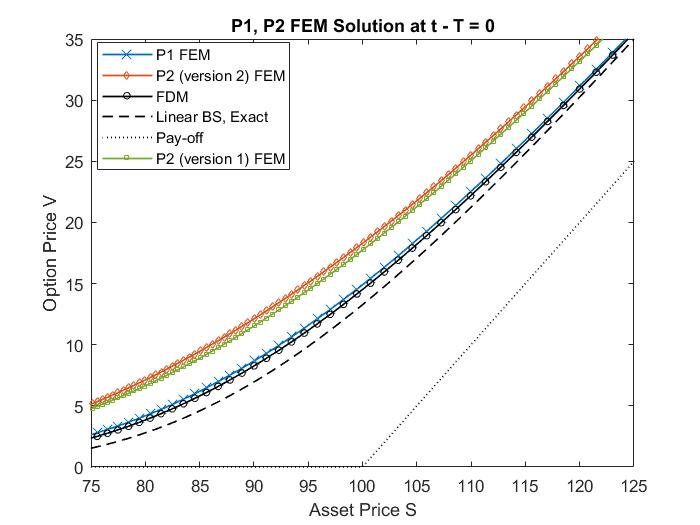}
\end{center}
\caption{Solution of Leland's model at 1 year towards expiration, with $r = 0.1$, $\sigma = 0.2$, $K = 100$, $c = 0.01$, $\delta t = 0.01$, giving $Le \approx 0.4$. {\bf Left}: $h = 0.1$, $\Delta \tau = 0.001$ ($\Delta \tau /h =  0.01$, $\Delta \tau/h^2 = 0.1$); {\bf Right}: $h = 0.0125$, $\Delta \tau = 0.000125$ ($\Delta \tau /h =  0.01$, $\Delta \tau/h^2 = 0.8$).}
\label{fig:Le04expire}
\end{figure}

To have a clearer picture of the computed solutions, we show in Figure~\ref{fig:Le04expire} the FEM solutions at expiration and compare them with a finite-difference (FDM) solution and the exact solution of the linear case. In this case, the option price based on Leland's model is higher than that based on the linear Black-Scholes model. While the finite-difference method used here is insensitive to the choice of the discretization parameters, a large difference in the solutions is observed for the finite-element method. This dependency on the discretization parameters is stronger in the P2 FEM than in the P1 FEM, whose solution remains close to the FDM solution. We also observe that setting $\bar{M} = M$ in the P2 FEM (Version 2) does not result in a solution which is significantly different from Version 1 (with $M_1 \neq M$).

\begin{figure}[!h]
\begin{center}
\includegraphics[width=0.45\textwidth]{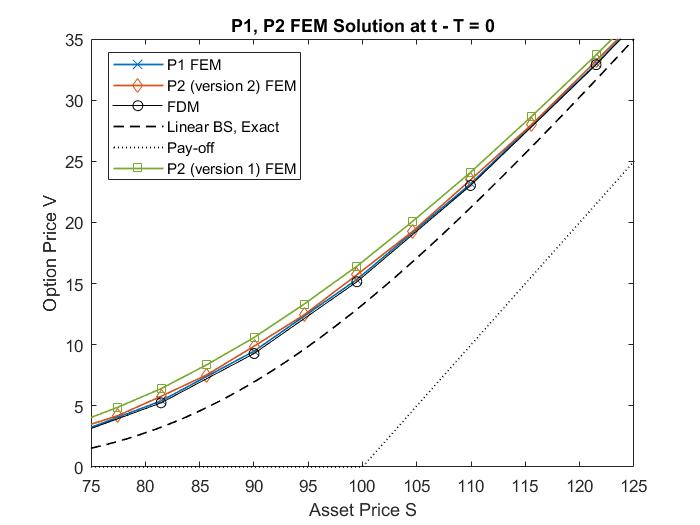}
\includegraphics[width=0.45\textwidth]{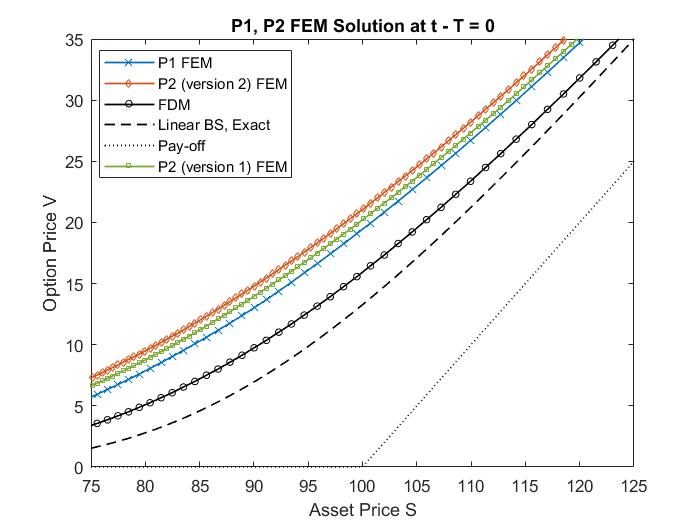}
\end{center}
\caption{Solution of  Leland's model at 1 year towards expiration, with $r = 0.1$, $\sigma = 0.2$, $K = 100$, $c = 0.02$, $\delta t = 0.01$, giving $Le \approx 0.8$;  {\bf Left}: $h = 0.1$, $\Delta \tau = 0.001$ ($\Delta \tau/h = 0.01$, $\Delta \tau/h^2 = 0.1$); {\bf Right}: $h = 0.0125$, $\Delta \tau = 0.000125$ ($\Delta \tau/h = 0.01$, $\Delta \tau/h^2 = 0.8$).} 
\label{fig:Le08expire}
\end{figure}

A second test in this class of problems uses the same  parameters, except for $c$, which is now set to $0.02$ (higher round-trip of the transaction cost). This setting corresponds to $Le = 0.8$. Numerical solutions over the time $t \in [0,1]$ (not shown in this paper) do not indicate any instability, as in the previous case. A closer observation of the solution at expiration also shows a higher price obtained by Leland's model  than that by the linear Black-Scholes model (see Figure~\ref{fig:Le08expire}). The FEM results also show the strong-dependency on the numerical parameters, leading to a higher computed option price. Furthermore, Version 1 and Version 2 do not result in significantly different solutions. 

\paragraph{The $Le > 1$ case.} For this class of problem, we set $c = 0.03$, while keeping the same values for the other parameters as in the $Le < 1$ cases. This results in $Le = 1.2$. Since Version 1 and 2 do not lead to a significant difference in the solutions,  we therefore will only implement and show results using Version 1 in the next numerical tests. 

\begin{figure}[!h]
\begin{center}
\includegraphics[width=0.45\textwidth]{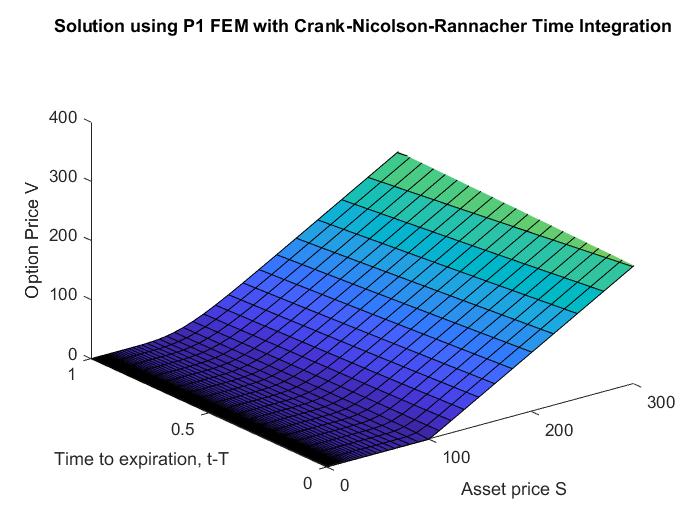}
\includegraphics[width=0.45\textwidth]{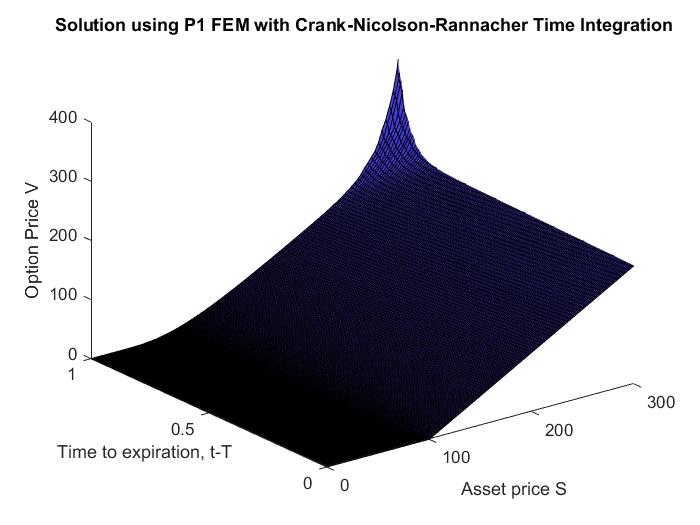}
\end{center}
\caption{P1-FEM solutions of  Leland's model, with $r = 0.1$, $\sigma = 0.2$, $K = 100$, $c = 0.03$, $\delta t = 0.01$, giving $Le \approx 1.2$. {\bf Left}: $h = 0.1$, $\Delta \tau = 0.001$ ($\Delta \tau/h = 0.01$, $\Delta \tau/h^2 = 0.1$); {\bf Right}: $h = 0.0125$, $\Delta \tau = 0.000125$ ($\Delta \tau/h = 0.01$, $\Delta \tau/h^2 = 0.8$).}
\label{fig:Le1.2P1}
\end{figure}

\begin{figure}[!h]
\begin{center}
\includegraphics[width=0.45\textwidth]{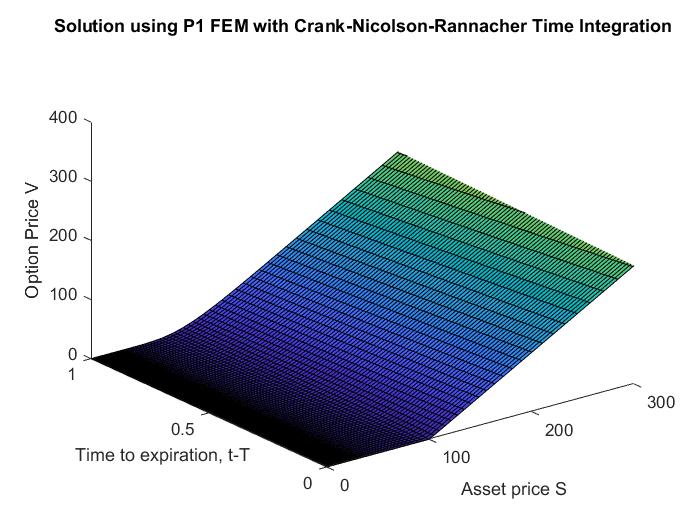}
\includegraphics[width=0.45\textwidth]{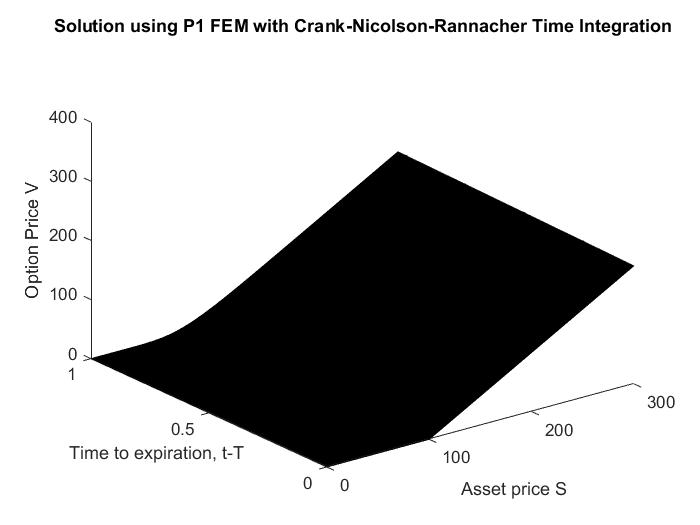}
\end{center}
\caption{P1-FEM solution  of  Leland's model, with $r = 0.1$, $\sigma = 0.2$, $K = 100$, $c = 0.03$, $\delta t = 0.01$, giving $Le \approx 1.2$. {\bf Left}: $h = 0.05$, $\Delta \tau = 0.00025$ ($\Delta \tau/h = 0.005$, $\Delta \tau/h^2 = 0.1$); {\bf Right}: $h = 0.025$, $\Delta \tau = 0.0000625$ ($\Delta \tau/h = 0.0025$, $\Delta \tau/h^2 = 0.1$). }
\label{fig:Le1.2P1stable}
\end{figure}

Figure~\ref{fig:Le1.2P1} presents numerical results using the P1 FEM, with $\Delta t/h^2 = 0.1$ and $0.8$ (see the detail in the figure caption). For this test problem, instability is observed in the numerical result with $\Delta t/h^2 = 0.8$, as $t$ gets closer to 0 (the initial time). This instability can be suppressed by setting a smaller temporal step such that $\Delta t /h = 0.005$, while keeping the same ratio $\Delta \tau/h^2$, as demonstrated by the numerical results in Figure~\ref{fig:Le1.2P1stable}.

\begin{figure}[!h]
\begin{center}
\includegraphics[width=0.45\textwidth]{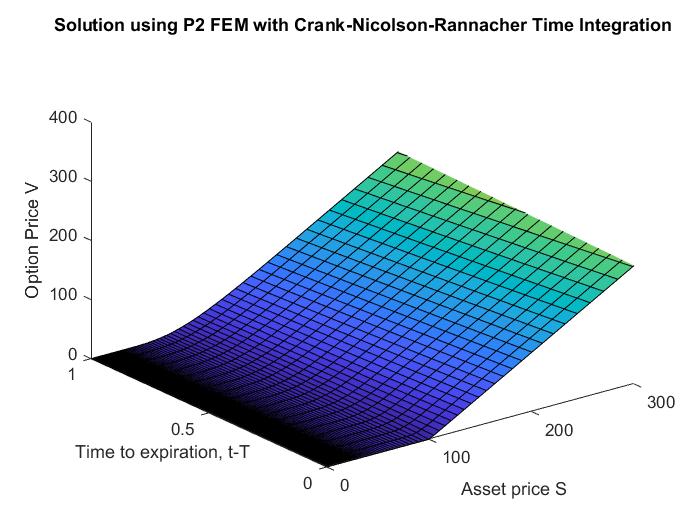}
\includegraphics[width=0.45\textwidth]{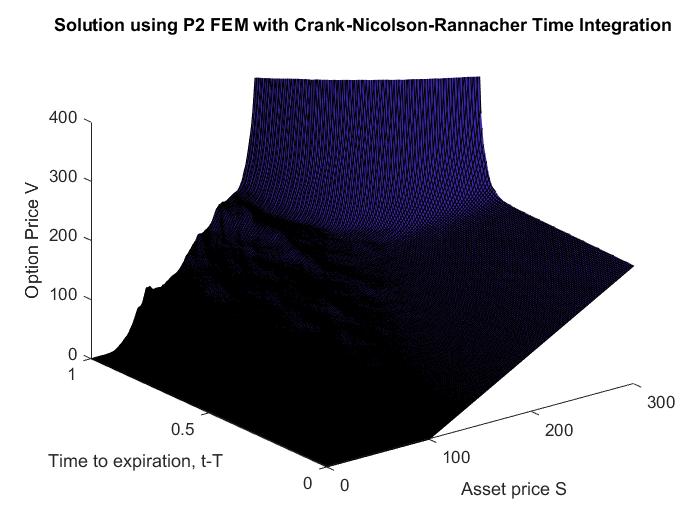}
\end{center}
\caption{P2-FEM solution of the Leland's model, with $r = 0.1$, $\sigma = 0.2$, $K = 100$, $c = 0.03$, $\delta t = 0.01$, giving $Le \approx 1.2$. {\bf Left}: $h = 0.1$, $\Delta \tau = 0.001$ ($\Delta \tau/h = 0.01$, $\Delta \tau/h^2 = 0.1$); {\bf Right}: $h = 0.0125$, $\Delta \tau = 0.000125$ ($\Delta \tau/h = 0.01$, $\Delta \tau/h^2 = 0.8$). }
\label{fig:Le1.2P2}
\end{figure}

\begin{figure}[!h]
\begin{center}
\includegraphics[width=0.45\textwidth]{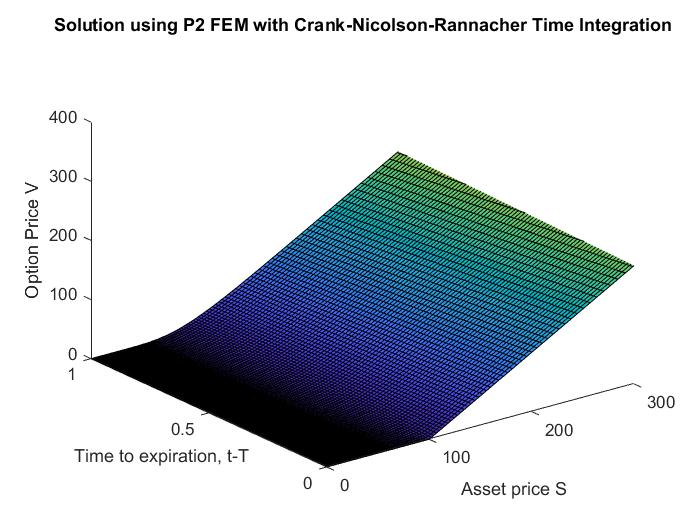}
\includegraphics[width=0.45\textwidth]{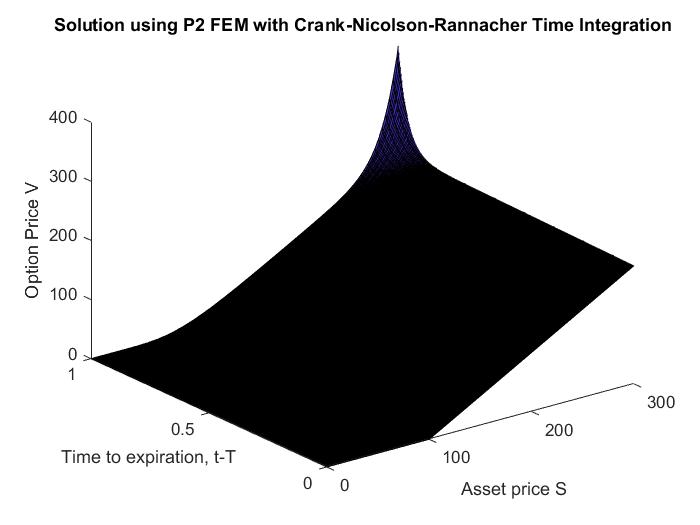}
\end{center}
\caption{P2-FEM solution  of the Leland's model, with $r = 0.1$, $\sigma = 0.2$, $K = 100$, $c = 0.03$, $\delta t = 0.01$, giving $Le \approx 1.2$. {\bf Left}: $h = 0.05$, $\Delta \tau = 0.00025$ ($\Delta \tau/h = 0.005$, $\Delta \tau/h^2 = 0.1$); {\bf Right}: $h = 0.025$, $\Delta \tau = 0.0000625$ ($\Delta \tau/h = 0.0025$, $\Delta \tau/h^2 = 0.1$). }
\label{fig:Le1.2P2stable}
\end{figure}

Using the above-mentioned test setting, we compute the solution with the P2 FEM. Numerical solutions with $\Delta \tau/h^2 = 0.1$ do not indicate any instability, both on the course and fine spatial-temporal mesh; see Figures~\ref{fig:Le1.2P2} and~\ref{fig:Le1.2P2stable}. Instability is observed in the numerical solutions with $\Delta \tau/h^2 = 0.8$, with spurious oscillation on the coarse spatial-temporal mesh. Such an instability can however still be controlled by decreasing the spatial and temporal step, even though in the current test, it is not fully eliminated.

\section{Conclusion} \label{conclusion}

In this work, several finite element-based models for approximations of the solutions of the Leland's model are built in combination with the Crank-Nicolson-type temporal scheme. It is demonstrated through several numerical examples that stable and accurate solutions can be obtained by these models  by  controlling the  spatial finite element size and the temporal steps. These numerical results compare favorably with those computed by finite difference schemes.   Our finite element models  can be used as an effective alternative for numerical solutions of the Leland model including the standard Black-Scholes model.
  
\bibliography{main}
\bibliographystyle{plain}

\end{document}